\documentclass[10pt,prd,twocolumn,floatfix,tightenlines,showpacs,showkeys,preprintnumbers,altaffilletter, nofootinbib, superscriptaddress, longbibliography]{revtex4-1}
\usepackage{graphicx,mathtools,tabu}
\usepackage{epstopdf}
\usepackage{amsmath}
\usepackage{amsfonts}
\usepackage[colorlinks=true,citecolor=red,linkcolor=blue,breaklinks=true]{hyperref}
\usepackage[dvipsnames]{xcolor}
\usepackage{accents}
\usepackage[figure, table]{hypcap}
\usepackage{amssymb}
\usepackage{appendix}
\usepackage{comment}
\usepackage{bbold}
\usepackage{color}
\usepackage{slashed}
\usepackage{subfigure}
\usepackage{setspace}
\usepackage{footnote}
\usepackage{multirow}
\usepackage{braket}
\usepackage[normalem]{ulem}
\usepackage[utf8]{inputenc}
\usepackage{mathrsfs}
\usepackage{longtable}
\usepackage{enumitem}

\date{\today}

\begin{document}
\preprint{}
\title{
Supernova fast flavor conversions in 1+1D: Influence of mu-tau neutrinos\\
}

\author{Francesco Capozzi}
\email{fcapozzi@ific.uv.es}
\affiliation{Instituto de F\'isica Corpuscular, Universidad de Valencia \& CSIC, Edificio Institutos de Investigaci\'on, Calle Catedr\'atico Jos\'e Beltr\'an 2, 46980 Paterna, Spain}

\author{Madhurima Chakraborty}
\email{madhu176121012@iitg.ac.in}
\affiliation{Indian Institute of Technology,
Guwahati, Assam-781039, India}

\author{Sovan Chakraborty}
\email{sovan@iitg.ac.in}
\affiliation{Indian Institute of Technology, 
Guwahati, Assam-781039, India}

\author{Manibrata Sen}
\email{manibrata@mpi-hd.mpg.de}
\affiliation{Max-Planck-Institut f{\"u}r Kernphysik, Saupfercheckweg 1, 69117 Heidelberg, Germany}

\begin{abstract}


In the dense supernova environment, neutrinos can undergo fast flavor conversions which depend on the large neutrino-neutrino interaction strength. It has been recently shown that both their presence and outcome can be affected when passing from the commonly used three neutrino species approach to the more general one with six species. Here, we build up on a previous work performed on this topic and perform a numerical simulation of flavor evolution in both space and time, assuming six neutrino species. We find that the results presented in our previous work remain qualitatively the same even for flavor evolution in space and time. This emphasizes the need for going beyond the simplistic approximation with three species when studying fast flavor conversions.

\end{abstract}

\maketitle

\section{Introduction}
\label{sec:Introduction}
Neutrino flavor conversions in the context of extremely dense astrophysical environments remain perhaps one of the biggest unsolved theoretical problems in neutrino physics. The main reason is that in such circumstances the neutrino-neutrino interaction potential is not negligible, as it usually is everywhere else, thus making the evolution deeply nonlinear. This gives rise to the phenomena known as collective oscillations, where the neutrinos having different energies undergo flavor conversion in a coherent manner \cite{Pantaleone:1994ns,Duan:2010bg,Mirizzi:2015eza,Chakraborty:2016yeg,Horiuchi:2017sku,Tamborra:2020cul,Capozzi:2022slf}. Depending on the timescale required for the development of these self-induced flavor conversions, they are classified as slow or fast. The growth rate of the slow modes is given by $\sqrt{\mu_0\omega_{\rm vac} } (\sim 10^2-10^3$ km$^{-1})$, where $\mu_0=\sqrt{2} G_F n_{\nu}$ is neutrino-neutrino interaction strength, whereas $\omega_{\rm vac}=\frac{\Delta m^2}{2E}$ is the much smaller vacuum oscillation frequencies. In comparison to this, the growth of fast modes is dependent only on $\mu_0$ which can be as large as $\sim 10^5$ km$^{-1}$ in the dense core.   

A large number of studies \cite{Hannestad:2006nj,Duan:2007mv,PhysRevD.83.105022,Fogli_2007,Johns:2017oky, Sawyer:2015dsa, Chakraborty:2016lct,Dasgupta:2016dbv,Airen:2018nvp,Izaguirre:2016gsx,Li:2021vqj,Padilla-Gay:2021haz,Wu:2021uvt,Richers:2021nbx,Richers:2021xtf,Martin:2019gxb,Abbar:2018beu,Johns:2020qsk,Sigl:2021tmj,Xiong:2021dex,Abbar:2021lmm,DelfanAzari:2019tez,Nagakura:2019sig,Nagakura:2021suv,Abbar:2020fcl,Nagakura:2021hyb,Stapleford:2019yqg,Capozzi:2018rzl,Wu:2017drk,Zaizen:2019ufj,Raffelt:2007yz,Esteban-Pretel:2007jwl,Duan:2005cp} have been carried out in the last two decades in order to investigate both the presence and outcome of collective oscillations. However, the corresponding system of partial differential equations has never been solved in its entire form, but only using some simplifying assumptions. For instance, it is usually assumed that there are only three neutrino species: $\nu_e\,\bar{\nu}_e$, and $\nu_x$, where one is considering $\nu_x=\nu_\mu=\nu_\tau=\bar{\nu}_\mu=\bar{\nu}_\tau$ i.e. all the heavy lepton species have identical fluxes. This has important consequences for fast conversions, since in this case the necessary and sufficient condition for their occurrence  \cite{Sawyer:2005jk,Sawyer:2008zs,Sawyer:2015dsa,Chakraborty:2016lct,Dasgupta:2016dbv,Izaguirre:2016gsx,Capozzi:2017gqd, Dighe:2017sur, Dasgupta:2017oko,Dasgupta:2018ulw,Abbar:2018shq,Azari:2019jvr,Johns:2019izj,Glas:2019ijo,Shalgar:2019qwg,Bhattacharyya:2020dhu,Bhattacharyya:2020jpj,Bhattacharyya:2021klg,Morinaga:2021vmc, Dasgupta:2021gfs,Bhattacharyya:2022eed,Zaizen:2021wwl} is the presence of a crossing {\it only} in the electron lepton number angular distribution. This means that in some directions the flux of $\nu_e$ is greater than that of $\bar{\nu}_e$ and vice versa in the other directions. 
In the context of fast flavor conversions, the first investigations going beyond three neutrino species  have been performed in Refs. \cite{Chakraborty:2019wxe,Capozzi:2020kge,Shalgar:2021wlj}. It has been shown with both numerical simulations and linear stability analyses, that even small differences in the angular distributions of $\nu_{\mu,\tau}$ and $\bar{\nu}_{\mu,\tau}$ can either create new instabilities or erase the ones present in the three species case. Furthermore, in Ref. \cite{Shalgar:2021wlj}, it has been pointed out that, even considering the same flavor content in the three and six neutrino species cases, the flavor conversion probabilities obtained as output of numerical simulations can have appreciable differences. However, the previous conclusions have been obtained assuming only the evolution in time, whereas spatial homogeneity has been imposed.

In this work, focusing again on fast conversions, we extend the studies performed in Refs. \cite{Chakraborty:2019wxe,Capozzi:2020kge,Shalgar:2021wlj} by considering the dependence of flavor evolution on the spatial dimension as well. In particular, we consider the same neutrino angular distributions that we considered in Ref. \cite{Capozzi:2020kge}. By extending the fast flavor conversion three-flavor analysis to $1+1$ dimensions, i.e., involving time and spatial evolution, we demonstrate the robustness of the results provided in Ref. \cite{Capozzi:2020kge}. We further discuss a consistent way of comparing the analysis in the three-species case with the analysis in the six-species case in a manner in which the net neutrino content is similar in two cases. 

The structure of the paper is as follows. In Sec. \ref{sec:Equations of motion}, we introduce the framework of the system and explain the equations of motion. Then, we elaborate the four toy examples taken into account for the analysis of the system with six species in Sec. \ref{sec:Toy Angular Distributions: 3 flavor analysis}. This is followed by Sec. \ref{sec:Linearized Regime} where we talk about the evolution of the system in the linearized regime and solve the dispersion relations for the four toy cases. Then we discuss the full nonlinear evolution considering one space and time dimension in Sec. \ref{sec:Full space-time evolution}. Finally, in Sec. \ref{sec:Comparison}, we present a comparison of the 1+1-dimensional analysis between the three and six species cases, taking into consideration the same flavor content for both scenarios.

\section{Framework : Equations of motion}
\label{sec:Equations of motion}
The equations of motion describing the spatial and temporal evolution of the neutrino density matrices $\rho_{{\bf p}, {\bf x},t}$ for momentum ${\bf p}$ at position ${\bf x}$ and time $t$ can be written in the form ~\cite{Sigl:1992fn} 
\begin{equation}
i\left(\partial_t + {\bf v}_{\bf p} \cdot \nabla_{\bf x}\right) \rho_{{\bf p}, {\bf x},t} 
= [H_{{\bf p}, {\bf x},t}, \rho_{{\bf p}, {\bf x},t}] 
\,\ ,
\label{eq:eom1}
\end{equation}
where $H_{{\bf p}, {\bf x},t}$ is the Hamiltonian of the system which consists of three parts, i.e., vacuum term, Mikheyev-Smirnov-Wolfenstein potential and the neutrino-neutrino interaction terms given by:
\begin{equation}
H_{\rm vac} = \Delta m^2/2\,E
\label{eq:eom1.1}
\end{equation}
\begin{equation}
H_{\rm mat} = \sqrt{2}G_F n_\alpha
\label{eq:eom1.2}
\end{equation}
\begin{equation}
H_{\nu \nu} = \mu_0\,\int d^3{\bf q}/(2\pi)^3 (1-{\bf v}_{\bf p}\cdot {\bf v}_{\bf q})({\rho}_{{\bf q}, {\bf x},t}-{\bar\rho}_{{\bf q}, {\bf x},t})
\,\ 
\label{eq:eom1.3}
\end{equation}
Here, $n_\alpha$ denotes the charged lepton density ($\alpha$ denotes the flavor), and the neutrino-neutrino interaction strength is given by $\mu_0=\sqrt{2} G_F n_{\nu}$, where $n_{\nu}$ is the total background neutrino density and $G_F$ is the Fermi constant. The diagonal elements of $\rho_{{\bf p}, {\bf x},t}$ represent the occupation numbers for each neutrino flavor, whereas the off-diagonal elements encode phase information related to flavor conversions. For the evolution of the antineutrinos, an equation similar to \ref{eq:eom1} holds with $H_{\rm vac}$ replaced by -$H_{\rm vac}$. 

In our previous work \cite{Capozzi:2020kge}, we studied only the time evolution of a system with six neutrino species, but here our aim is to take into account one space and time dimension, i.e., 1+1 dimensions. Furthermore, while studying space and time evolution, we neglect both H$_{\rm vac}$ (since its role is just to provide a numerical seed for the development of fast conversions) and H$_{\rm mat}$~\cite{Dasgupta:2018ulw, Abbar:2017pkh}.

In the three neutrino species approach, fast conversions are triggered when there is a crossing in the electron lepton number (ELN), which is defined as~\cite{Izaguirre:2016gsx}
\begin{equation}
G^e_{\bf v} = \sqrt{2} G_F \int_{0}^{\infty}\frac{dE\,E^2}{2 \pi^2}\left[\rho_{ee}(E,{\bf v})-\bar{\rho}_{ee}(E,{\bf v})\right] \,\,.
\label{eq:eln}
\end{equation}
Considering six neutrino species, as we do in this work, we can define also a muon lepton number (MuLN) and tau lepton number (TauLN),
\begin{eqnarray}
G^\mu_{\bf v} &=&\sqrt{2} G_F \int_{0}^{\infty}\frac{dE\,E^2}{2 \pi^2}\left[\rho_{\mu\mu}(E,{\bf v})-\bar{\rho}_{\mu\mu}(E,{\bf v})\right] \,\,,\\
G^\tau_{\bf v} &=&\sqrt{2} G_F \int_{0}^{\infty}\frac{dE\,E^2}{2 \pi^2}\left[\rho_{\tau\tau}(E,{\bf v})-\bar{\rho}_{\tau\tau}(E,{\bf v})\right] \,\,.
\label{eq:mutau-ln}
\end{eqnarray}
In this case, fast conversions occur when a crossing is present in one of the following three quantities
\begin{eqnarray}
 G^{e\mu}_{\bf v}&=&G^e_{\bf v}-G^\mu_{\bf v} \,\,,\nonumber\\
 G^{e\tau}_{\bf v}&=&G^e_{\bf v}-G^\tau_{\bf v}\,,\nonumber\\
 G^{\mu\tau}_{\bf v}&=&G^\mu_{\bf v}-G^\tau_{\bf v}\,.
 \label{eq:Gv}
 \end{eqnarray}
 The recent 2D simulations~\cite{Bollig:2017lki} provide support to this possibility. It suggests that the temperatures in the accretion phase are high enough for the creation of muons through the pair production from electrons which in turn can create $\nu_\mu$ and $\bar{\nu}_\mu$ by the means of $\beta$ processes. This leads to an asymmetry between the $\mu$ neutrinos and antineutrinos. However, the high mass value of the $\tau$ lepton restricts the production of $\nu_\tau$ and $\bar{\nu}_\tau$ through similar processes, but still there can be a small asymmetry between them because of their different scattering cross sections with nucleons.
 
A crossing in $G^{\alpha\beta}_{\bf v}$ will first lead to an exponential growth of the off diagonal elements of $\rho_{\alpha\beta}$, which will then propagate to the other density matrices $\rho_{\alpha_1\beta_1}$ ($\alpha_1\beta_1\ne\alpha\beta$). In other words, the growth in any one of the three sectors can trigger the growth in the others. This is in contrast with the three neutrino species scenario where a crossing in the ELN is considered to be the only requirement for whether the fast oscillations will occur or not.

\section{Toy Angular Distributions:\newline three-flavor analysis}
\label{sec:Toy Angular Distributions: 3 flavor analysis}
To study fast flavor oscillations in the six species scenario, we consider four toy examples (the same as in Ref. \cite{Capozzi:2020kge}). The angular distributions as a function of $v=\cos{\theta}$ are given by the expression
\begin{equation}
   \rho_{\alpha\alpha}=\frac{1}{2\pi}\Big[\frac{1+\Delta_{\alpha}}{1-v_{min}^{\alpha}} \mathcal{H}(\cos{\theta}-v_{min})+h \mathcal{H}(-\cos{\theta}+v_{min})\Big] 
\end{equation}
where, $\alpha=e,\bar{e},\mu,\bar{\mu},\tau,\bar{\tau}$ and the parameters $\Delta_{\alpha},v_{min}^{\alpha}$, and h for four different cases are mentioned in Tables \ref{tab:table-1} and \ref{tab:table-2}. Here, $\mathcal{H}$ is the Heaviside Theta function. Our parametrization takes into account backward velocity modes, implying that there are neutrinos going in the backward direction, i.e., $-1<v <1$. 
\begin{table}[h!!!!]
  \begin{center}
 \begin{tabular}{ |p{0.8cm}|p{0.8cm}|p{0.8cm}|p{0.8cm}|} 
  \hline 
 $\alpha$ & $v_{min}^{\alpha}$ & $\Delta_{\alpha}$ & h\\
 \hline
 e & -1.00 & 0.80 & 0\\
 \hline
 $\bar{e}$ & -0.60 & 0.70 & 0\\
 \hline
  $\mu$ & -0.80 & 0.10 & 0\\
 \hline
   $\bar{\mu}$ & -0.70 & 0.45 & 0\\
 \hline
  $\tau$ & -0.80 & 0.10 & 0\\
 \hline
  $\bar{\tau}$ & -0.70 & 0.45 & 0\\
 \hline
  \end{tabular}
  \quad
  \begin{tabular}{ |p{0.8cm}|p{0.8cm}|p{0.8cm}|p{0.8cm}|} 
  \hline 
 $\alpha$ & $v_{min}^{\alpha}$ & $\Delta_{\alpha}$ & h\\
 \hline
 e & -1.00 & 0.80 & 0\\
 \hline
 $\bar{e}$ & -0.60 & 0.70 & 0\\
 \hline
  $\mu$ & -0.80 & 0.30 & 0\\
 \hline
  $\bar{\mu}$ & -0.70 & 0.15 & 0\\
 \hline
  $\tau$ & -0.80 & 0.30 & 0\\
 \hline
  $\bar{\tau}$ & -0.70 & 0.15 & 0\\
 \hline
  \end{tabular}
  \end{center}
     \caption{\label{tab:table-1}Parameter values for case 1 (left) and case 2 (right).}
  \end{table}
 \begin{table}[h!!!!]
  \begin{center}
 \begin{tabular}{ |p{0.8cm}|p{0.8cm}|p{0.8cm}|p{0.8cm}|} 
  \hline 
 $\alpha$ & $v_{min}^{\alpha}$ & $\Delta_{\alpha}$ & h\\
 \hline
 e & -1.00 & 0.90 & 0\\
 \hline
 $\bar{e}$ & -0.60 & 0.30 & 0\\
 \hline
  $\mu$ & -0.80 & 0.10 & 0\\
 \hline
  $\bar{\mu}$ & -0.70 & 0.50 & 0\\
 \hline
  $\tau$ & -0.80 & -0.20 & 0\\
 \hline
  $\bar{\tau}$ & -0.70 & -0.10 & 0\\
 \hline
  \end{tabular}
  \quad
  \begin{tabular}{ |p{0.8cm}|p{0.8cm}|p{0.8cm}|p{0.8cm}|} 
  \hline 
 $\alpha$ & $v_{min}^{\alpha}$ & $\Delta_{\alpha}$ & h\\
 \hline
   e & -0.30 & 0.60 & 0.00\\
 \hline
 $\bar{e}$ & 0.00 & 0.29 & 0.00\\
 \hline
  $\mu$ & -0.20 & 0.00 & 0.08\\
 \hline
  $\bar{\mu}$ & -0.10 & 0.20 & 0.02\\
 \hline
  $\tau$ & -0.20 & 0.10 & 0.08\\
 \hline
  $\bar{\tau}$ & -0.10 & 0.17 & 0.00\\
 \hline
  \end{tabular}
  \end{center}
      \caption{\label{tab:table-2}Parameter values for case 3 (left) and case 4 (right).}
  \end{table}

\begin{figure*}[!t]
\centering
\includegraphics[width=0.45\textwidth]{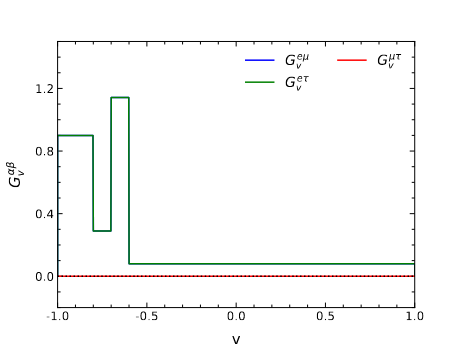}~
\includegraphics[width=0.45\textwidth]{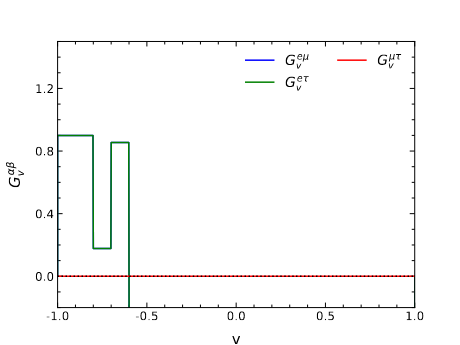}\\
\includegraphics[width=0.45\textwidth]{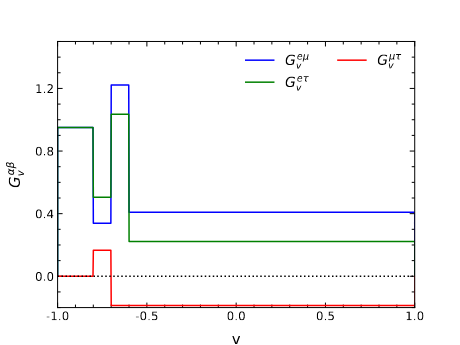}~
\includegraphics[width=0.45\textwidth]{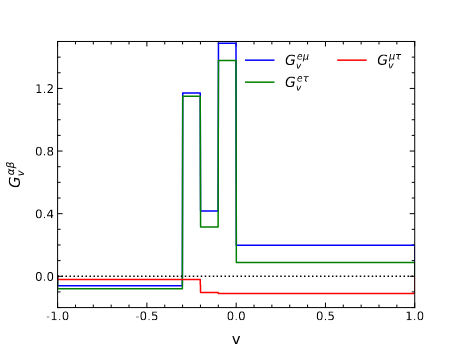}\\
\caption{The above panels show the effective lepton numbers for different flavors for the angular distributions of the four toy examples mentioned in the text. The upper left panel corresponds to case 1 and upper right panel is for case 2. The lower left and right panels represent the case 3 and case 4 respectively.}
\label{fig:angdist}
\end{figure*}

Figure \ref{fig:angdist} shows the differences between the angular distributions of different flavors given by Eq.~\ref{eq:Gv} for the four different cases mentioned above.

The upper left panel of Figure \ref{fig:angdist} represents the case 1 whose angular distributions are given by the parameter values in the left panel of Table \ref{tab:table-1}. It shows the difference of the lepton numbers in the case of three flavors. In this scenario, $G_\textbf{v}^{e\mu}$, $G_\textbf{v}^{\mu\tau}$ and $G_\textbf{v}^{e\tau}$ are shown by the blue, red and the green solid lines, respectively. Here, all three-flavor lepton number distributions (ELN, MuLN and TauLN) (not shown in the figure) have crossings, whereas the differences, i.e., G$_\textbf{v}^{e\mu}$, G$_\textbf{v}^{\mu\tau}$, and G$_\textbf{v}^{e\tau}$, do not have crossings. 

The upper right panel of Figure  \ref{fig:angdist} shows case 2,  whose angular distributions are given by the parameter values in the right panel of Table \ref{tab:table-1}. In this case, ELN has a crossing but MuLN and TauLN do not have. However, there is a crossing in G$_\textbf{v}^{e\mu}$ (blue solid line) and G$_\textbf{v}^{e\tau}$ (green solid line) but there is none in G$_\textbf{v}^{\mu\tau}$ (red solid line). 

Case 3 is represented by the lower left panel of Figure \ref{fig:angdist} and its angular distributions are given by the left panel of Table \ref{tab:table-2} . In this scenario, there is no crossing in ELN but it is present in MuLN and TauLN. Focusing on the differences, G$_\textbf{v}^{e\mu}$ (blue solid line) and G$_\textbf{v}^{e\tau}$ (green solid line) do not have a crossing whereas it is there in G$_\textbf{v}^{\mu\tau}$ (red solid line). 

Case 4 is given by the angular distributions with parameter values in the right panel of Table \ref{tab:table-2}. Here, there is a shallow crossing in the ELN in the forward direction and also there are crossings in the MuLN and TauLN. Unlike the other three cases, here the  differences G$_\textbf{v}^{e\mu}$ (blue solid line) and G$_\textbf{v}^{e\tau}$ (green solid line) have shallow backward crossings as shown in the lower right panel of Figure \ref{fig:angdist}. However, G$_\textbf{v}^{\mu\tau}$ (red solid line) does not have any crossing.

Taking these angular distributions into account, we study the temporal and spatial evolution of the system. First, we focus on the linearized regime by solving the dispersion relation and calculating the growth rates. Then, we move on to the nonlinear analysis where we numerically solve the evolution equation \ref{eq:eom1} in 1+1 dimensions. 

\section{Linearized Regime}
\label{sec:Linearized Regime}
The onset of the fast flavor conversions is studied through the method of linear stability analysis \cite{Izaguirre:2016gsx, Capozzi:2017gqd, Yi:2019hrp, Airen:2018nvp, Capozzi:2019lso, Chakraborty:2019wxe}. 
We linearize Eq. \ref{eq:eom1} at first order in the off-diagonal elements of the density matrices $S_{\textbf{v}}^{\alpha\beta}$, ($\alpha\ne\beta$), assuming the diagonal elements to be $O(1)$ \cite{Chakraborty:2019wxe}. This leads to the equations
\begin{equation}
  i\,v^\gamma\partial_\gamma S_{\textbf{v}}^{\alpha\beta}=
  \bigl(v^\gamma (\Lambda_{\gamma}^{\alpha\beta}+\Phi_{\gamma}^{\alpha\beta})\bigr) S_{\textbf{v}}^{\alpha\beta}
  -v^\gamma  \int \frac{d\textbf{v}'}{4\pi}\, v_\gamma'\,G_{\textbf{v}'}^{\alpha\beta}  S_{\textbf{v}'}^{\alpha\beta}\,,
  \label{eq:lineom}
\end{equation}
where, $\alpha\beta$ corresponds to the three sectors i.e., $e-\mu$, $e-\tau$ and $\mu-\tau$ respectively. Here, $\gamma=0,1,2,3$ and $\Lambda_{\gamma}^{\alpha\beta}$ is the charged lepton matter term and the corresponding current. Similarly, $\Phi_{\gamma}^{\alpha\beta}$ is the neutral lepton matter term and the corresponding current. Since we are neglecting the matter term, we take $\Lambda_{\gamma}^{\alpha\beta}=0$ and $\Phi_{\gamma}^{\alpha\beta}=\int \frac{d\textbf{v}}{4\pi}\, v_\gamma\,G_{\textbf{v}}^{\alpha\beta}$. 

\begin{figure*}[!t]
\includegraphics[width=0.3\textwidth]{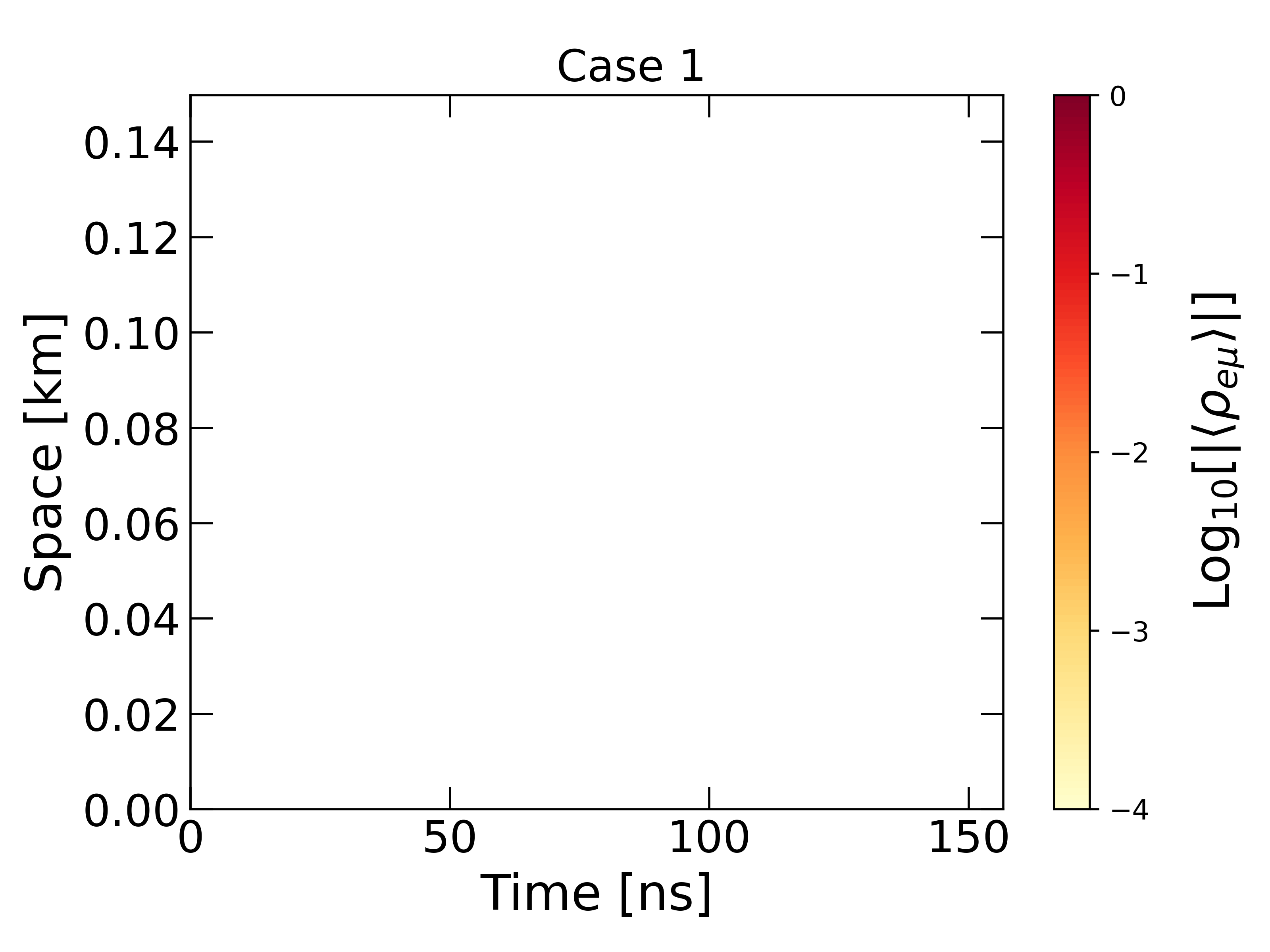}~\includegraphics[width=0.3\textwidth]{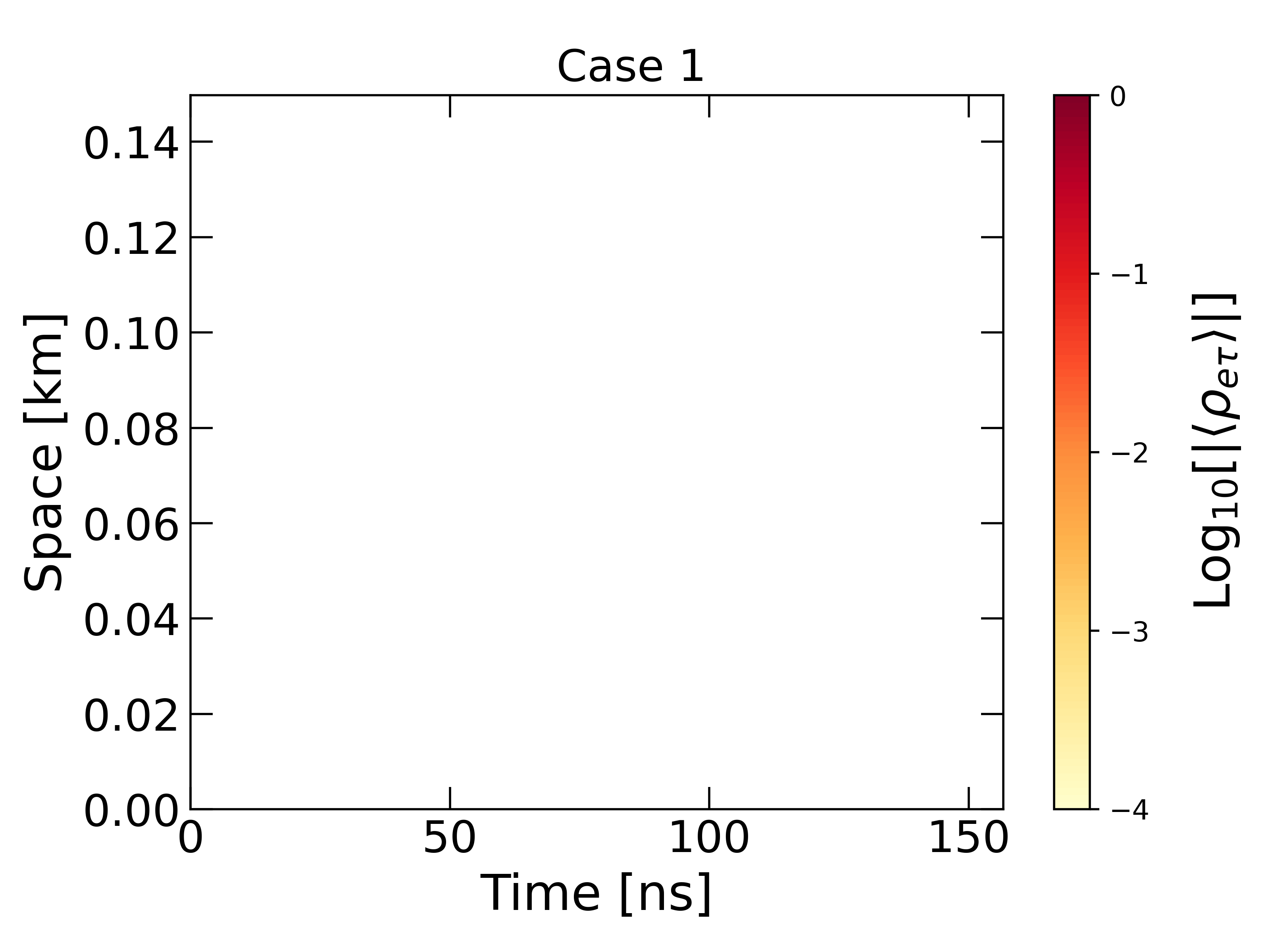}~
\includegraphics[width=0.3\textwidth]{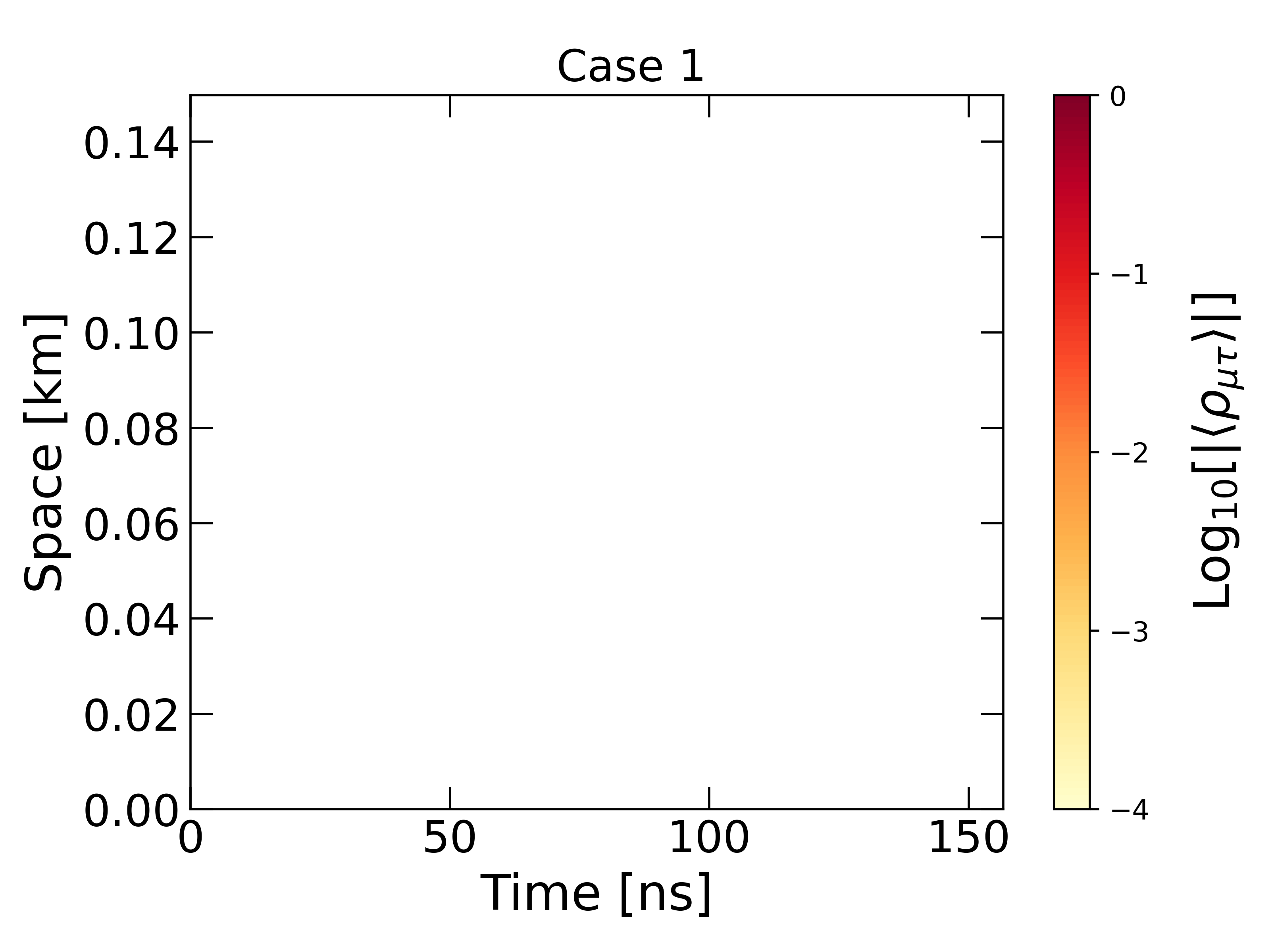}\\

\includegraphics[width=0.3\textwidth]{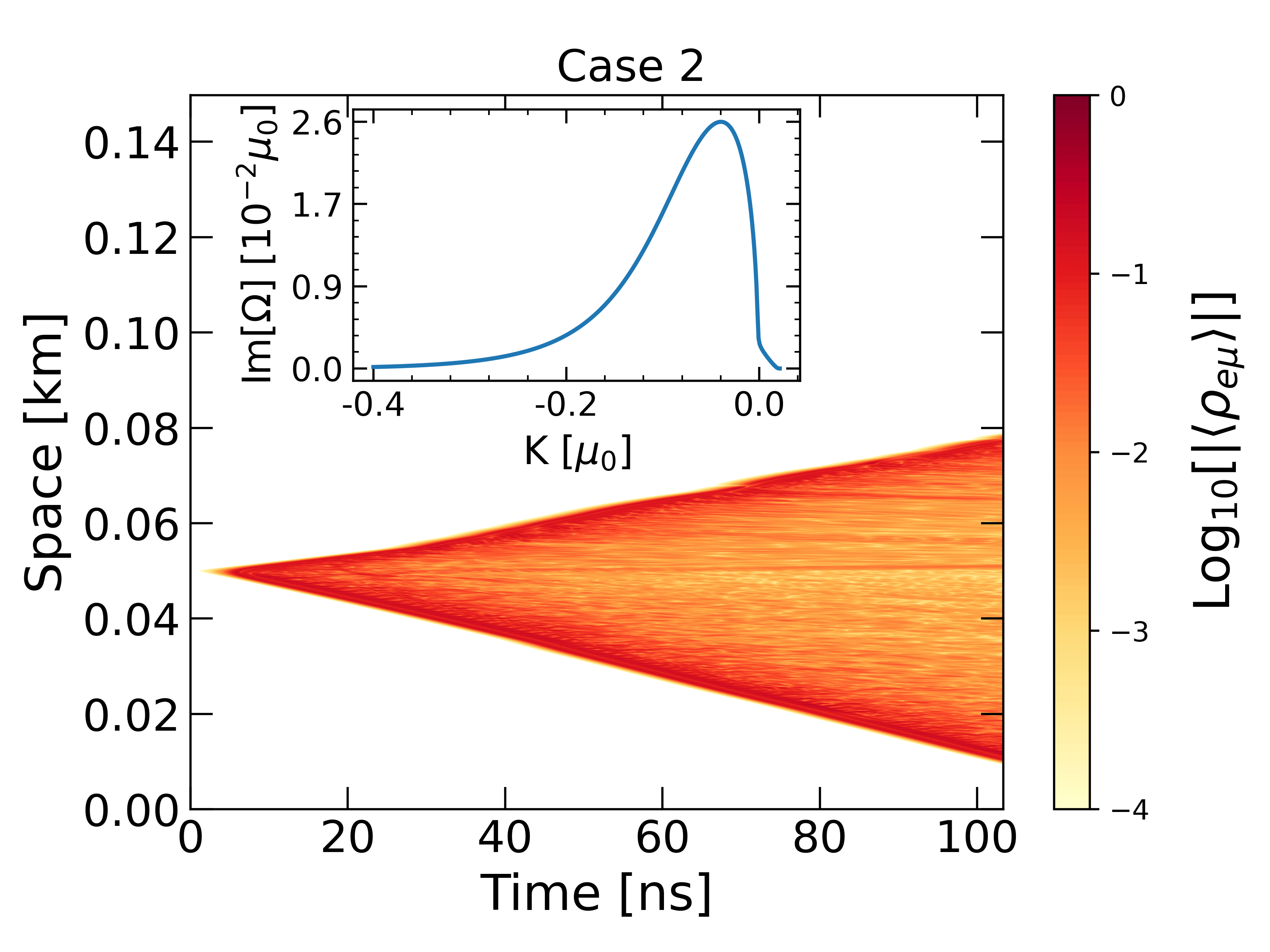}~\includegraphics[width=0.3\textwidth]{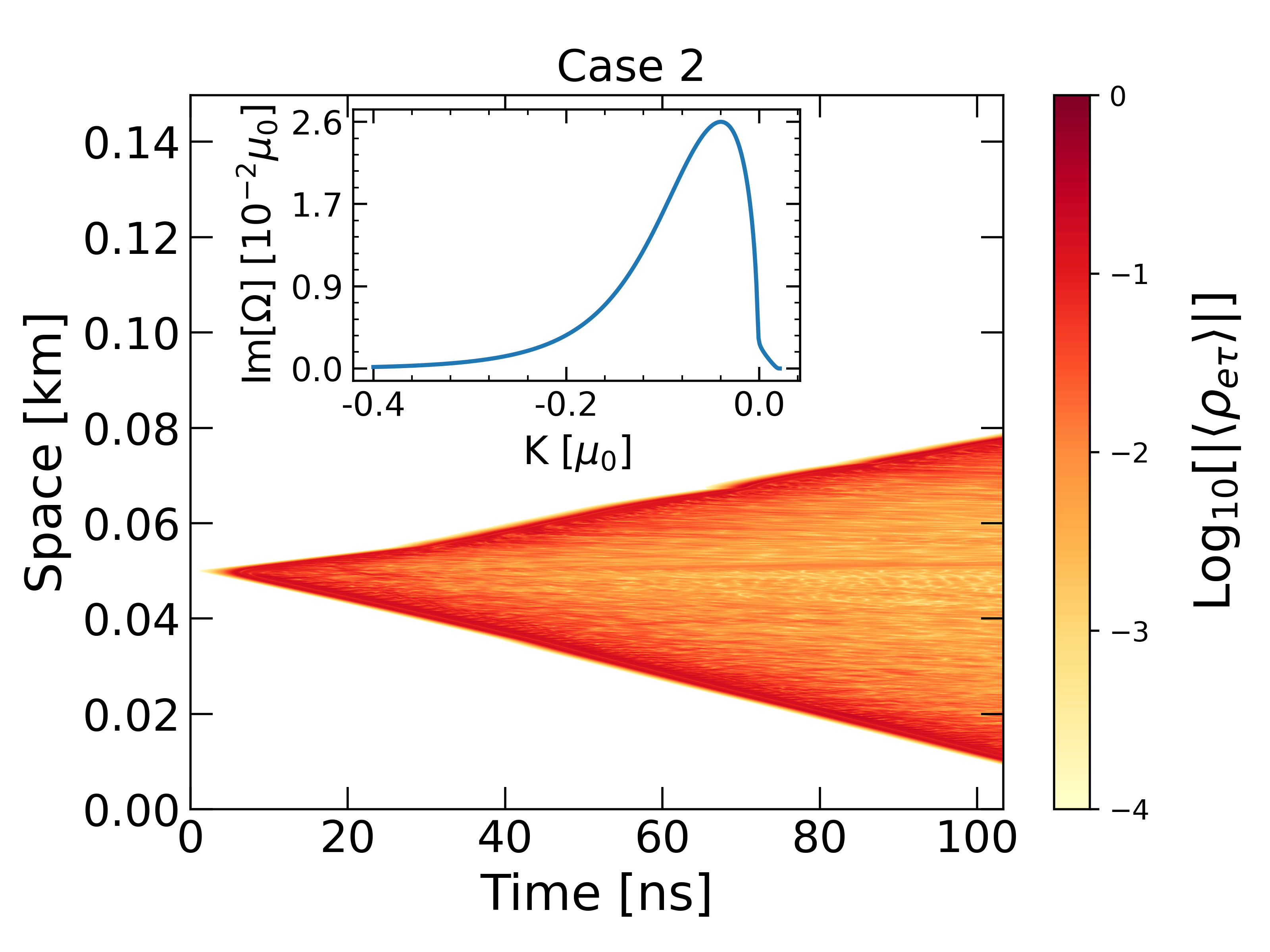}~
\includegraphics[width=0.3\textwidth]{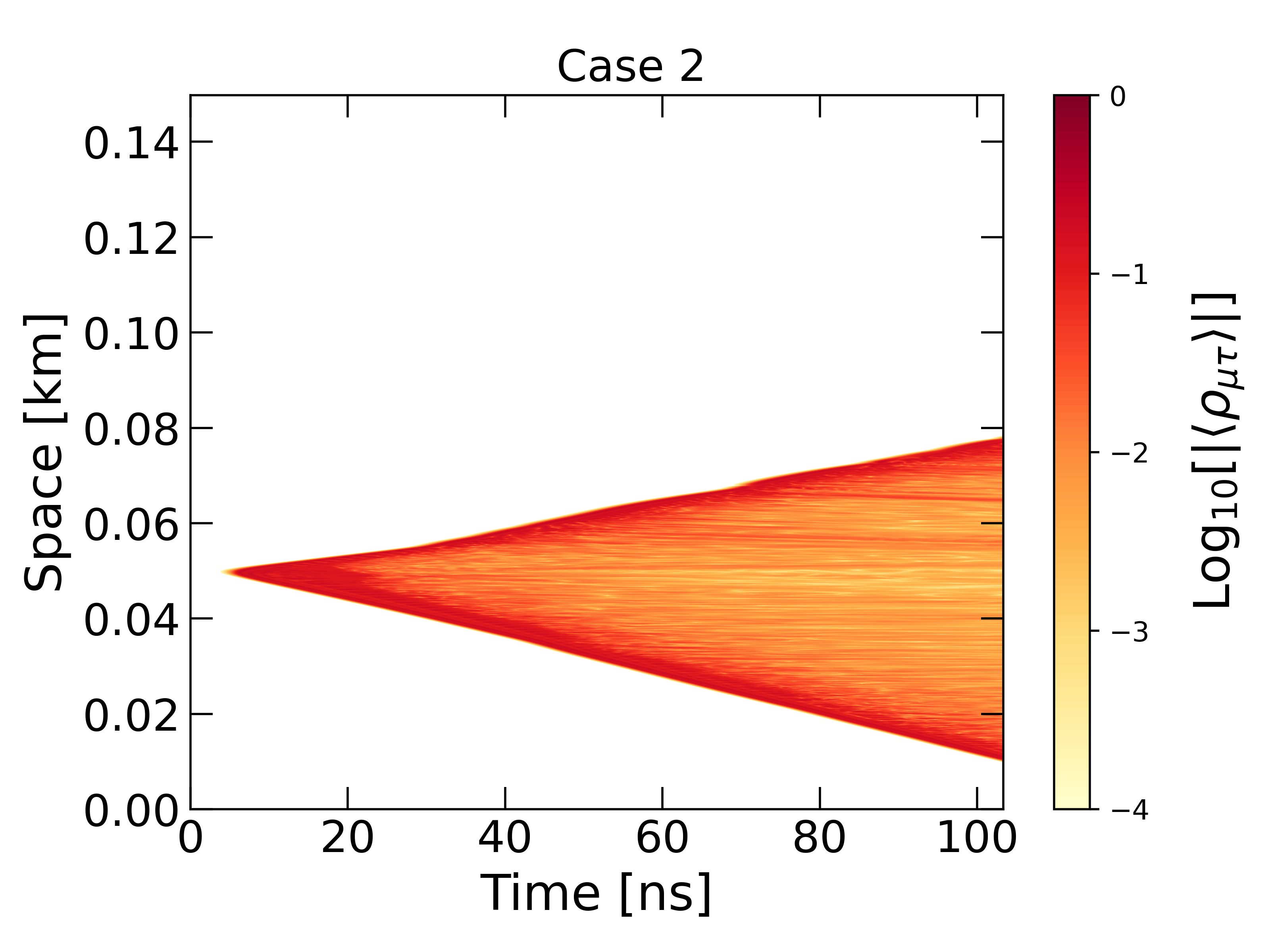}

\includegraphics[width=0.3\textwidth]{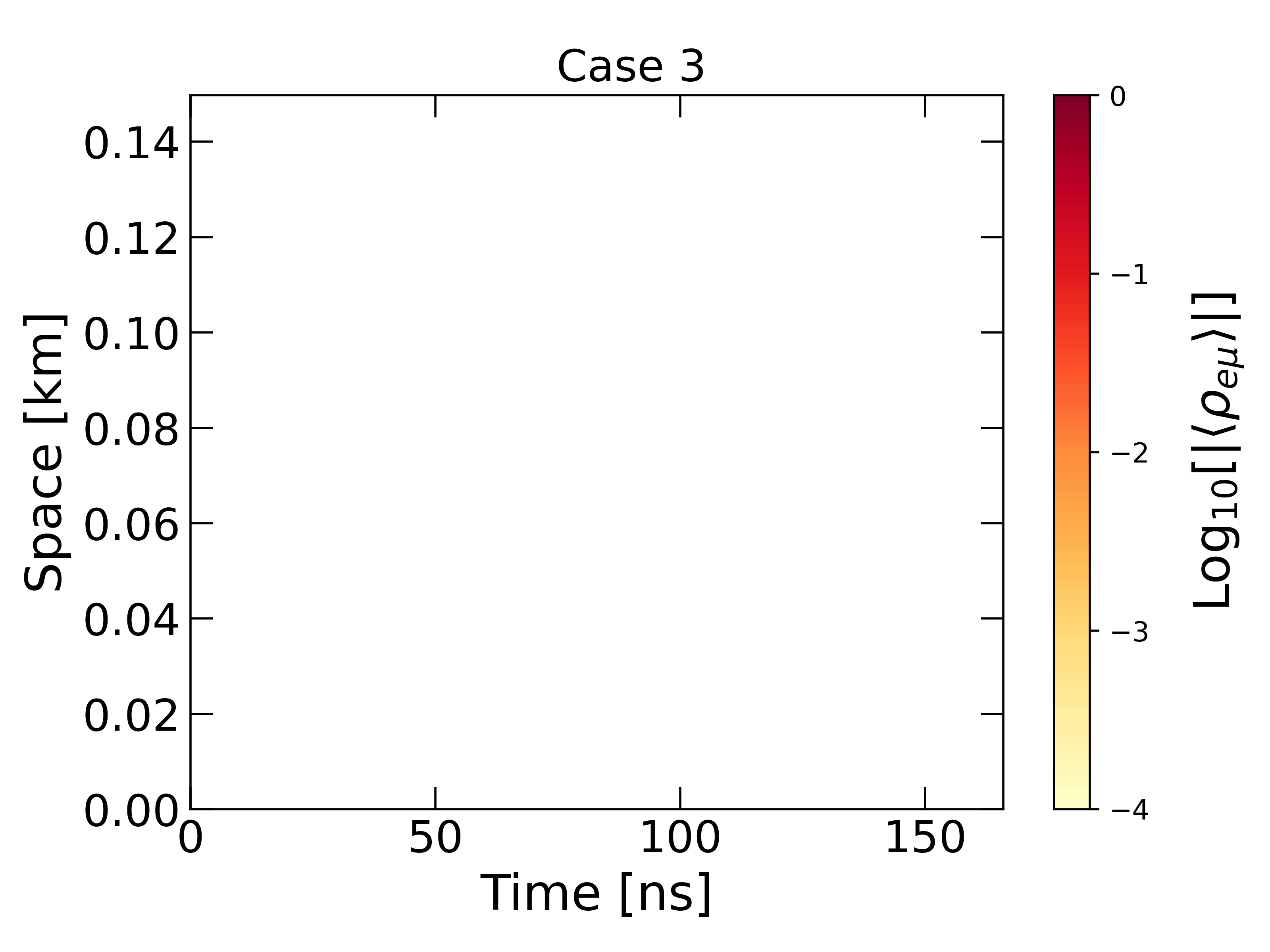}~\includegraphics[width=0.3\textwidth]{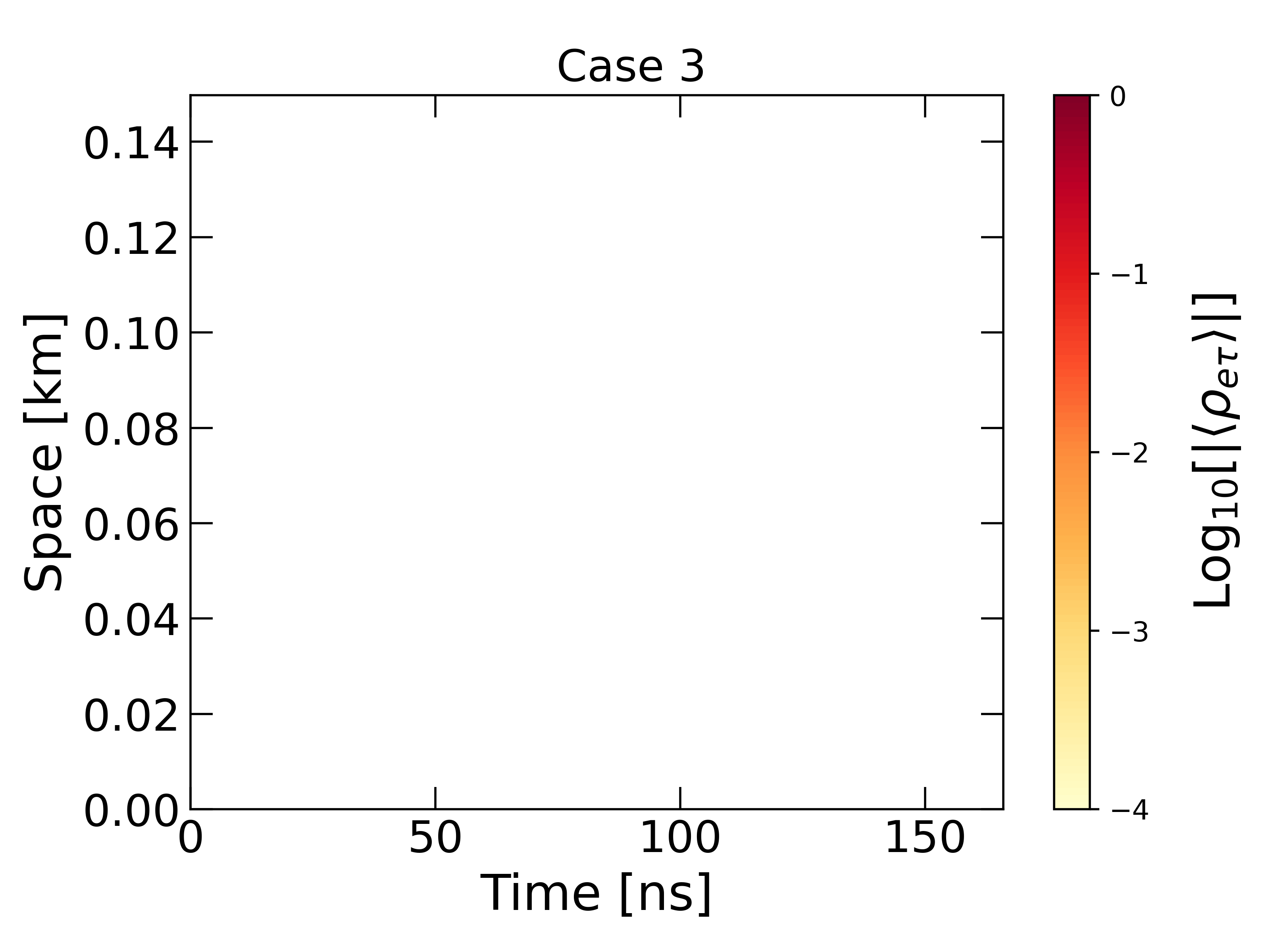}~
\includegraphics[width=0.3\textwidth]{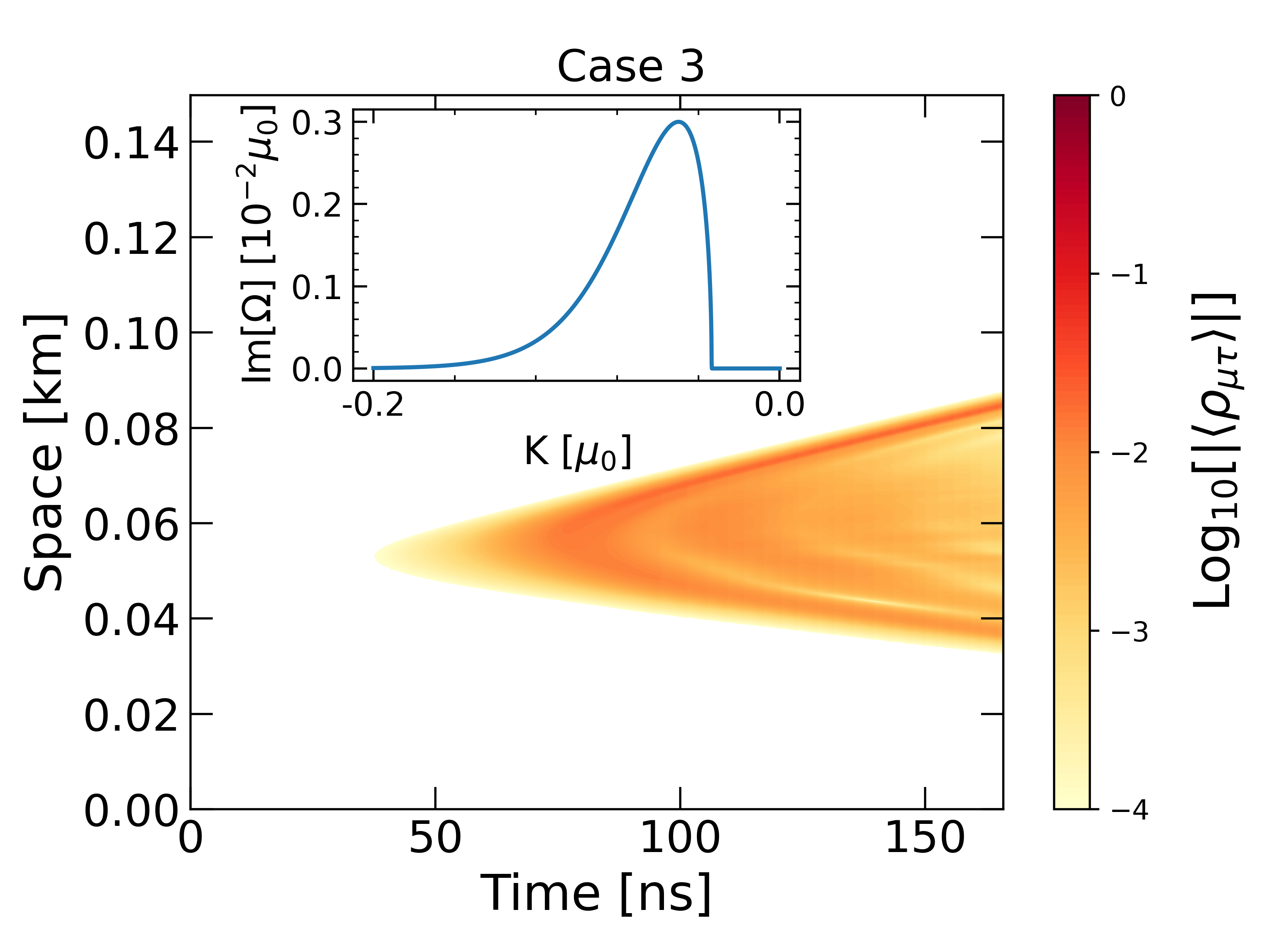}\\

\includegraphics[width=0.3\textwidth]{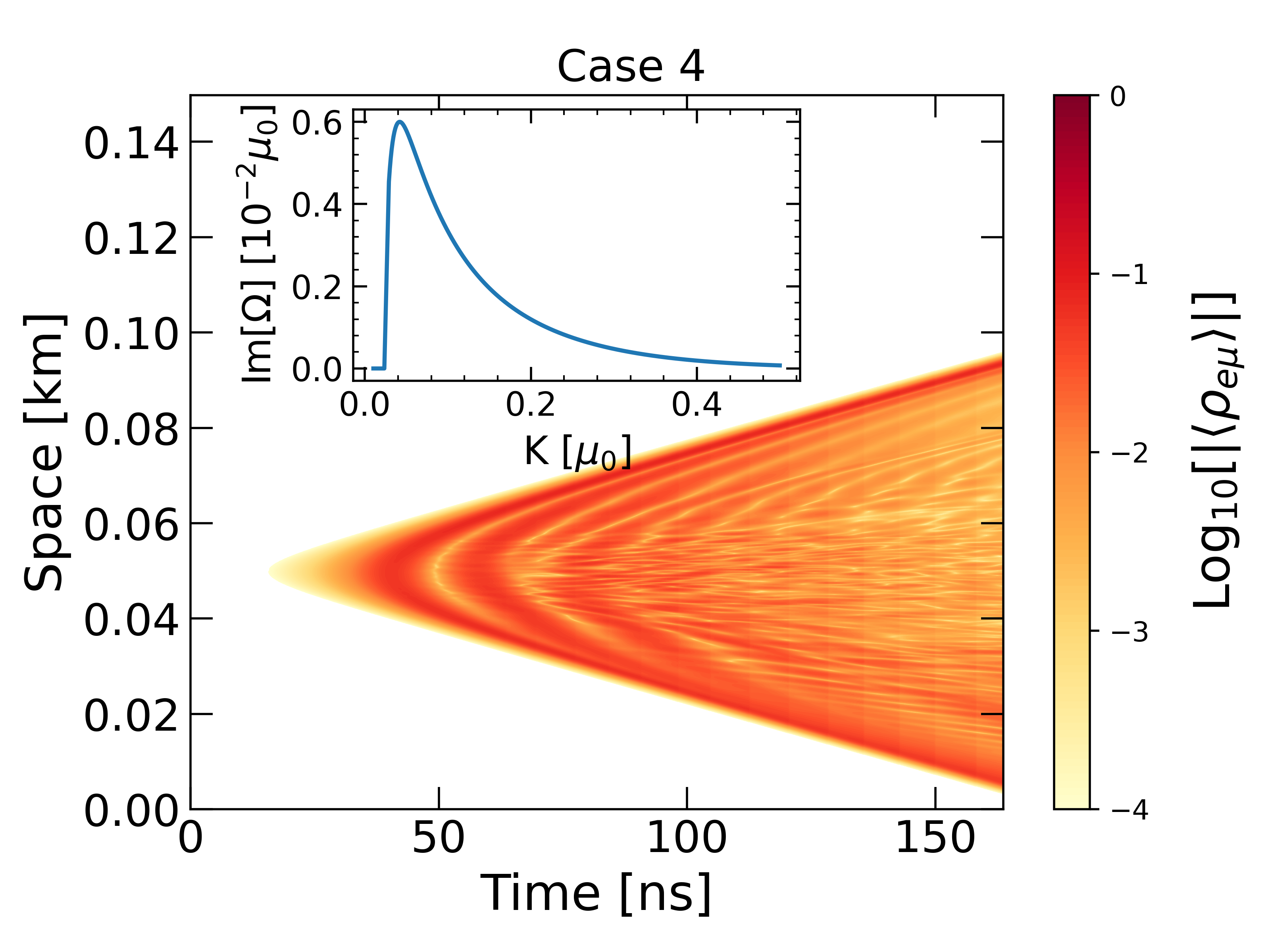}~\includegraphics[width=0.3\textwidth]{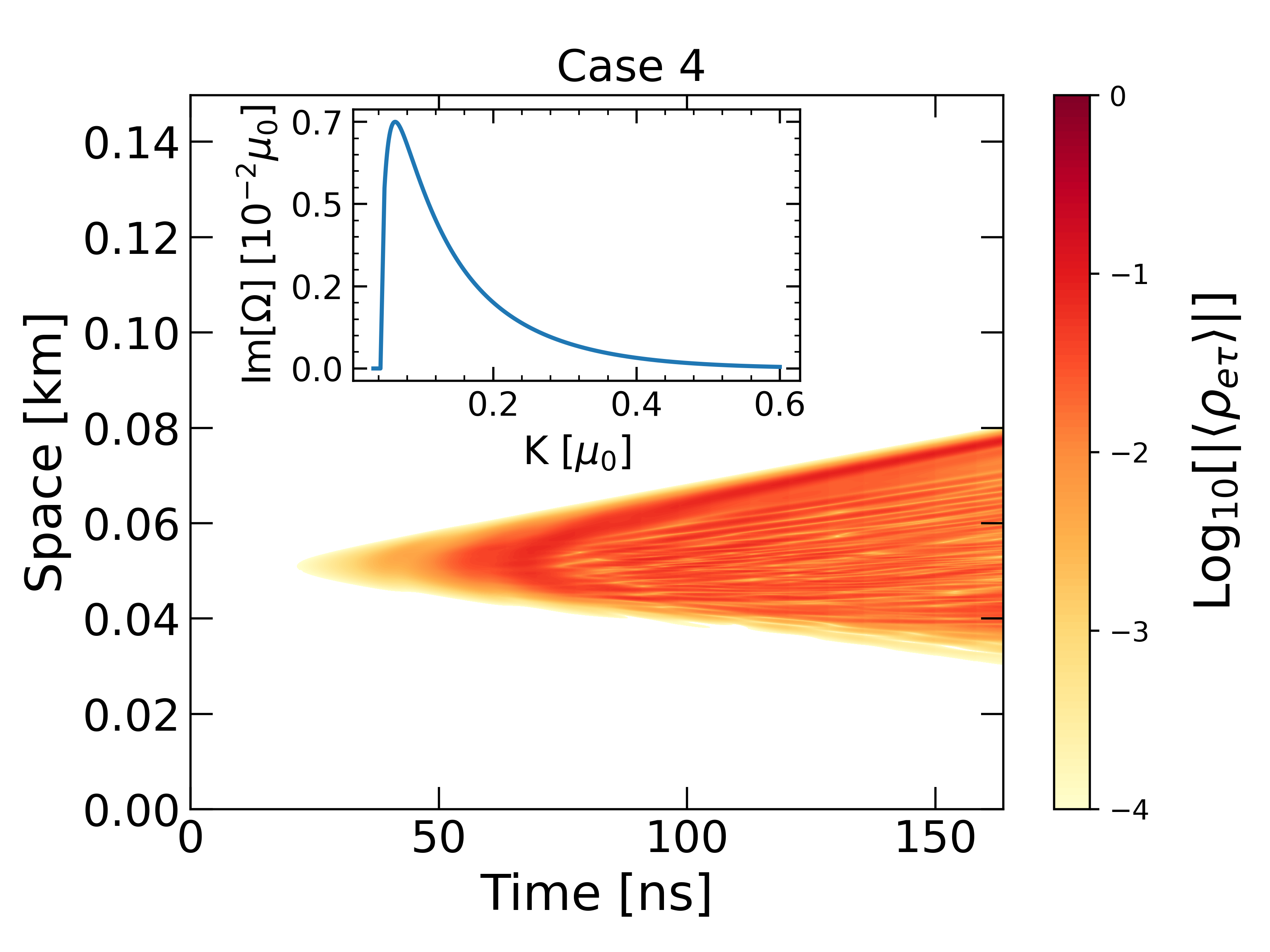}~
\includegraphics[width=0.3\textwidth]{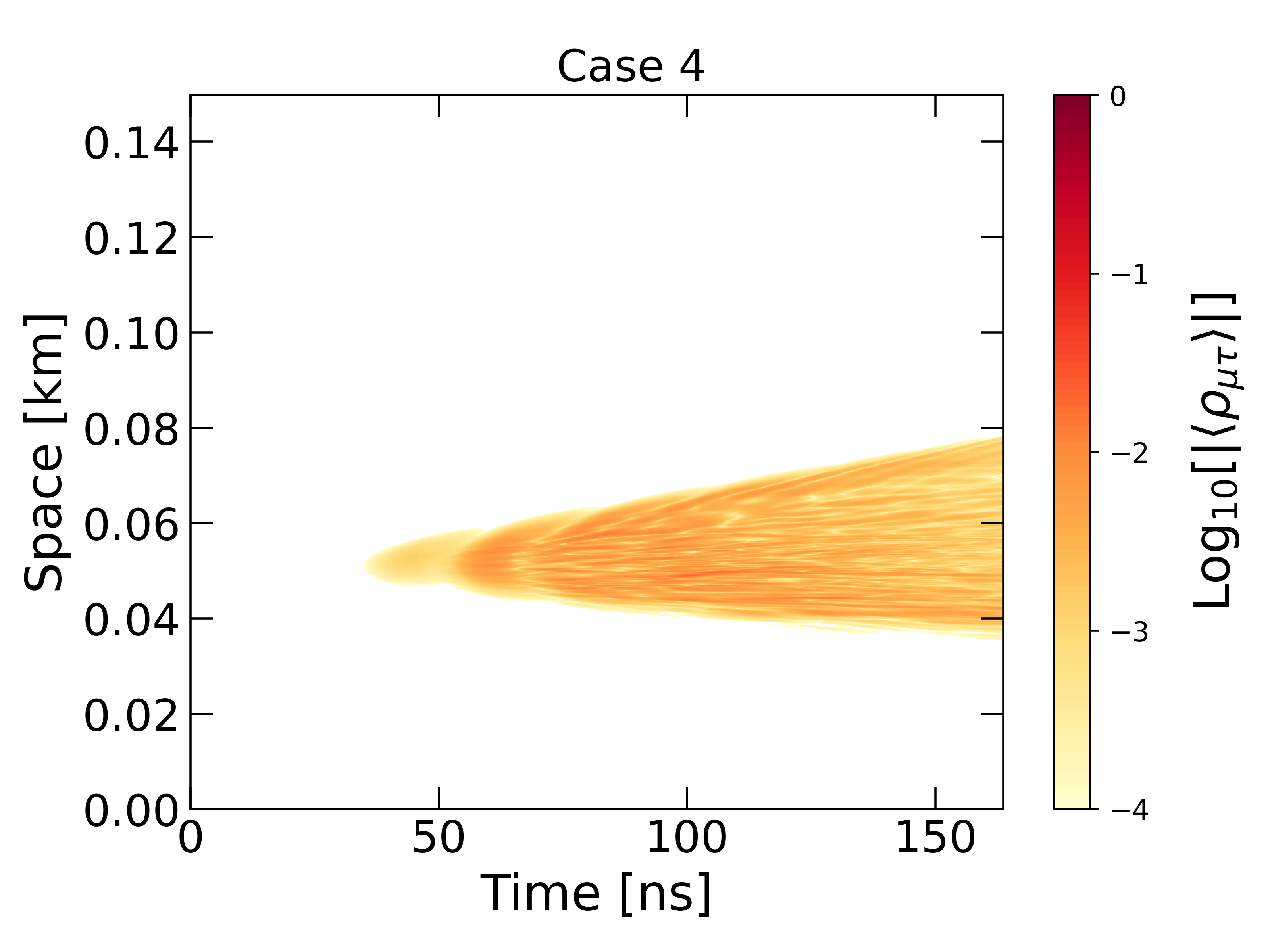}

\caption{Growth rate of flavor instability $(\log_{10}|\langle \rho_{\alpha\beta}\rangle|)$ in the three-flavor study for the four different cases. The rows (from top to bottom) correspond to cases 1,2,3, and 4, respectively. The left panel depicts the $e-\mu$ sector, while the middle and right panels depict the $e-\tau$ and the $\mu-\tau$ sectors respectively. The inset shows the linear growth rates plotted in units where $\sqrt{2}G_F n_\nu =1$.}
\label{fig:AllCase}
\end{figure*}

We then take the ansatz $S_{\textbf{v}}^{\alpha\beta}=Q_{\textbf{v}}^{\alpha\beta}\,e^{-i(\Omega t- \textbf{K}\cdot\textbf{x})}$. Substituting this back in Eq. \ref{eq:lineom}, we obtain the dispersion relation for a given choice of $\alpha\beta$, 
\begin{equation}
  {\rm det}\,[\Pi^{\gamma\delta}_{\alpha \beta}(\Omega,K)]=0\,,
  \label{eq:disprel}
\end{equation}
where the rank-2 polarization tensor $\Pi^{\gamma\delta}_{\alpha \beta}$ is given by:
\begin{equation}
  \Pi^{\gamma\delta}_{\alpha \beta}=\eta^{\gamma\delta}+\int \frac{d\textbf{v}}{4\pi}\,G_{\,\textbf{v}}^{\alpha\beta}\,
  \frac{v^\gamma v^\delta}{\Omega - (\textbf{K}-\Phi^{\alpha\beta}).\textbf{v}}\,
\end{equation}
where $\eta^{\gamma\delta}$ is the metric tensor and it is equal to diag(+1, -1, -1, -1). Note that the subscripts $\alpha, \beta$ of $\Pi$ denote the flavor of neutrinos and $\gamma, \delta$ are the spacetime indices. Here, there are three dispersion relations corresponding to the three sectors, i.e., $e-\mu$, $e-\tau$ and $\mu-\tau$. To investigate the presence of instabilities in the system, for each sector we solve Eq. \ref{eq:disprel} as a function of real values of K. If, we find Im$[\Omega(K)]\ne 0$, then we have an instability.


\section{Full space-time evolution}
\label{sec:Full space-time evolution}
In this section, we numerically solve the complete equations of motion Eq.\,\ref{eq:eom1} in one spatial dimension, $z$ and time $t$ for the four cases discussed above. The results of the simulations allow us to  compare the numerical growth rates with those expected from the stability analysis. 

For the numerical analysis, we employ the d03pff routine from the NAG library, which is built for solving a system of nonlinear convection-diffusion partial differential equations in one space dimension. The method of lines is employed to reduce the system of partial differential equations to a system of ordinary differential equations, and the resulting system is solved using a backward differentiation formula method. We use this routine with $10^3$ points in space and we discretize the angular variable $\mathbf{v}$ with 30 points. We fix $\mu_0=4\times 10^5$ km$^{-1}$ and we consider a spatial range $z\in [0,0.1]$ km. We set H$_{\rm vac}=0$, but we set the off-diagonal entries of $\rho_{\mathbf{p},z,t=0}$ equal to a Gaussian in space centered at $z=0.05$ km, with an amplitude of $10^{-9}$ and a width of $5\times 10^{-4}$ km. For boundary conditions, we assume that at $t=0$, the density matrix is empty at the edge of our spatial box. This means that neutrinos are only going outside of the box, whereas none is coming in. We use an absolute relative accuracy of $10^{-10}$ up to 30 ns for case 2 and up to 60 ns for cases 3 and 4, then, we switch to $10^{-6}$ in order to speed up calculations. For case 1, we always use $10^{-10}$. We have not imposed a maximum time step.

The results for the four cases, shown in Figure \ref{fig:AllCase}, demonstrate the growth rates of flavor instabilities in terms of the evolution of $\log_{10}|\langle \rho_{\alpha\beta}\rangle|$ in the $z-t$ plane, where $\langle \rho_{\alpha\beta}\rangle$ is angle averaged. The leftmost panels depict the evolution in the $e-\mu$ sector, while the middle panels show the same for the $e-\tau$ sector and the $\mu-\tau$ sector respectively. For the cases where we find a flavor instability, we show the corresponding linear growth rate in the inset plot. The growth rates have been obtained by solving the dispersion relation given by Eq. \ref{eq:disprel} taking into consideration the initial angular distributions of all the flavors.

We find that, as expected, there are no flavor instabilities for case 1 (top panel). This is consistent with the fact that the angular distributions in case 1 show no crossing in three flavors. In tandem with these results, we find null results for the growth rates using a stability analysis. For case 2 (second panel from top), there exists a crossing in the $e-\mu$ and the $e-\tau$ sectors. Consequently, the values of $\log_{10}|\langle \rho_{e\mu}\rangle|$ and $\log_{10}|\langle\rho_{e\tau}\rangle|$ start growing in space and time, indicating a flavor instability. The corresponding inset plots show the instabilities in the Im$\Omega-K$ plane in the linear regime. It can be seen that the instabilities are present for both positive and negative values of $K$. We also get a quantitative agreement among the amplitudes of the growth rates in the linear as well as nonlinear regimes. The growth-rate in the $\mu-\tau$ sector is driven by a combination of the couplings in the different sectors, and is purely driven by the nonlinearity of the problem. Hence, this growth rate is not captured by a stability analysis.

The toy spectra in case 3 (third panel from top) shows a crossing only in the $\mu-\tau$ sector. As a result, we find a flavor instability in the $\mu-\tau$ sector only. The linear stability analysis also obtains
imaginary values of $\Omega$ only in the $\mu-\tau$ sector and not in the other two. Furthermore, we note that the instabilities are present only for negative values of $K$.

For case 4 (bottom panel), we find that all three flavor sectors show a nonzero growth of the off-diagonal components of the density matrix. The $e-\mu$ and the $e-\tau$ sectors show a spectral crossing, and hence have a faster growth rate. The growth in the $\mu-\tau$ sector is completely a three-flavor artefact, and does not depend on any spectral crossing in that sector. The same is captured in a linear stability analysis, where there is an instability only in the $e-\mu$ and the $e-\tau$ sector, as shown in the inset. Note that here the instabilities are present only for the positive K. 

One interesting thing to note from the linear stability curves (insets of Figure \ref{fig:AllCase}) is that in the cases 3 and 4, there are no instabilities for the k = 0 mode. Here, `k' is the shifted wave vector in the corotating frame and is defined as k = K-$\Phi^{\alpha\beta}$, where $\Phi^{\alpha\beta}=\int \frac{d\textbf{v}}{4\pi}\, \textbf{v}\,G_{\textbf{v}}^{\alpha\beta}$ is the neutral lepton current term as defined in Sec. \ref{sec:Linearized Regime}. For the given spectra $G_{\textbf{v}}^{\alpha\beta}$ (shown in Figure \ref{fig:angdist}), in case 3, \textbf{$\Phi^{\mu\tau}$}$=-0.00528$ and similarly in case 4, \textbf {$\Phi^{e\mu}$}$=0.00371$ and \textbf {$\Phi^{e\tau}$}$=0.0298$ in units of $\mu_0$. From the insets of the lower two panels of Figure \ref{fig:AllCase}, one can see that Im($\Omega$) $=0$ corresponding to these values. In other words, no instabilities are present in cases 3 and 4 for K = $\Phi^{\alpha\beta}$, i.e., k = 0 mode. This also agrees with the fact that when we calculate the Im($\Omega$) following the formalism of Ref. \cite{Dasgupta:2018ulw} by finding the moments, we do not obtain any instability for these cases. This is because the moments formalism is based on the calculation of Im($\Omega$) for the case of k=0. On the other hand, for case 2, \textbf {$\Phi^{e\mu}=\Phi^{e\tau}$} $=-0.024$, corresponding to which a nonzero Im($\Omega$) exists (shown in the insets of second panel from top of Figure \ref{fig:AllCase}), although it is not the maximum value. Therefore, for this case, we obtain instability from the moments calculation.

\begin{figure*}[!t]
\includegraphics[width=0.3\textwidth]{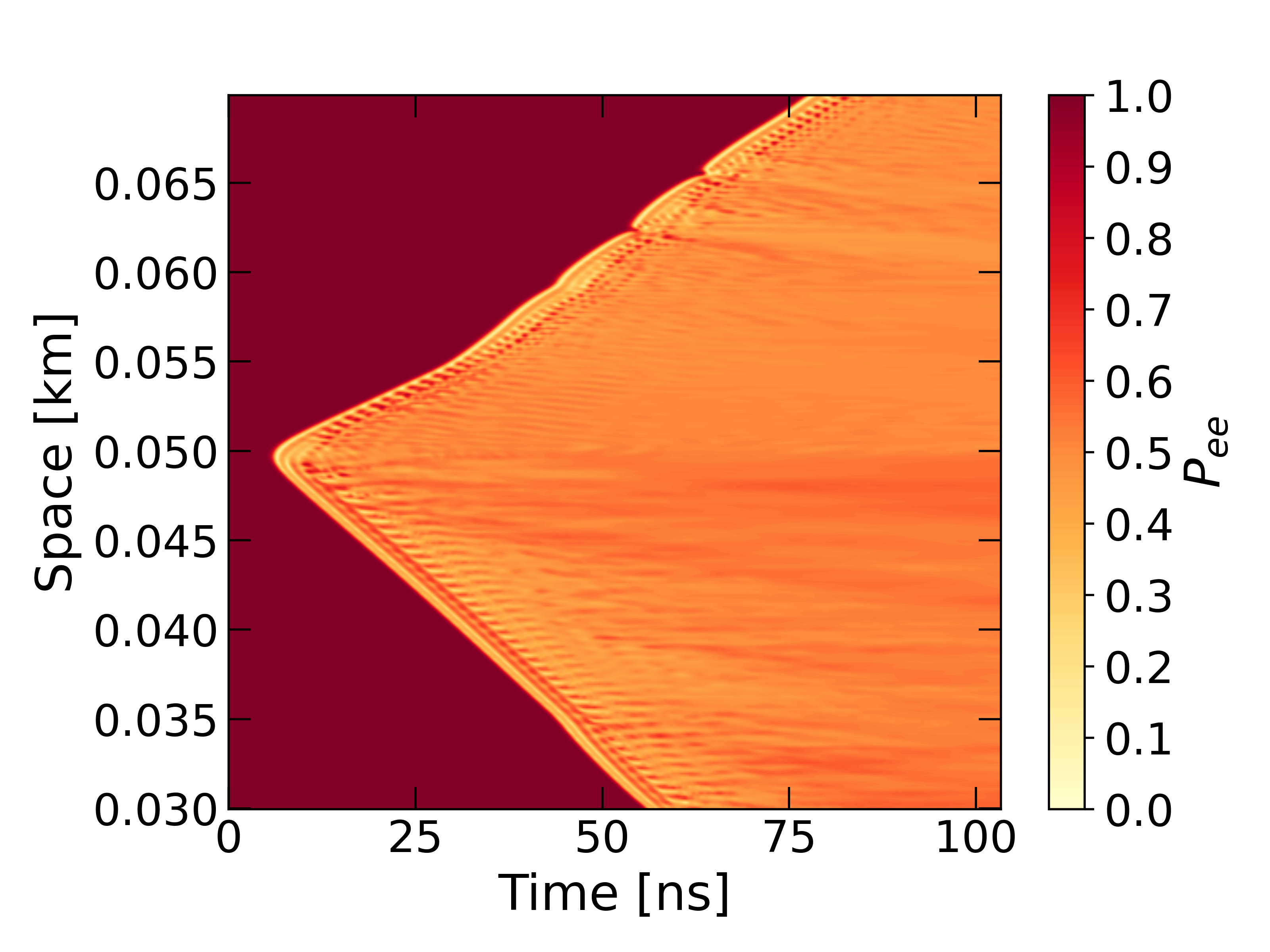}~\includegraphics[width=0.3\textwidth]{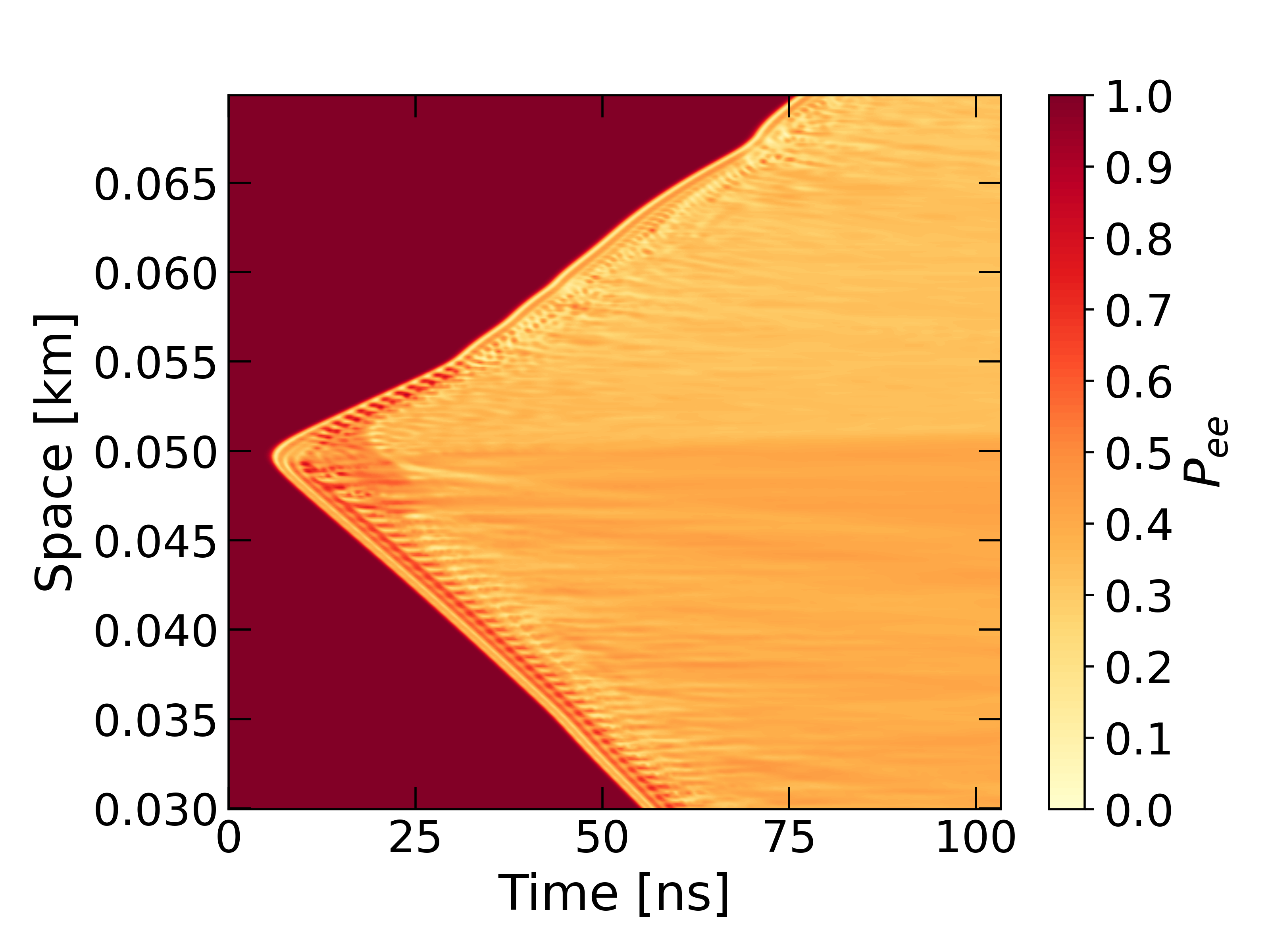}\\
\caption{Survival probability $P_{ee}$ obtained with three neutrino species (left panel) and six (right panel). In both panels the flavor evolution is obtained assuming $\rho_{\tau\tau}(z,t=0)=0$ and the same values of $\rho_{ee}(z,t=0)$, $\bar{\rho}_{ee}(z,t=0)$, $\rho_{\mu\mu}(z,t=0)$ and $\bar{\rho}_{\mu\mu}(z,t=0)$ taken from case 2. }
\label{fig:AllCase_Pee}
\end{figure*}

\section{Two- and three-flavor calculations with the same flavor content}
\label{sec:Comparison}

In this section we compare the survival probability $P_{ee}=(\langle\rho_{ee}(t,z)\rangle-\langle\rho_{\mu\mu}(0,z)\rangle)/(\langle\rho_{ee}(0,z)\rangle-\langle\rho_{\mu\mu}(0,z)\rangle)$ between the six and the three neutrino species approach, focusing only on case 2. Note that here $\langle \rho_{\alpha \alpha} \rangle$ is the angle sum. To make a fair comparison, we require that the total number of neutrinos at $t=0$ remains the same in both approaches. Indeed, since flavor evolution of fast conversions is dominated by the self-interaction term, which is proportional to the net neutrino number density, we believe that setting the initial neutrino number density to be the same is the only way to compare three species and six species analyses. As a result, we consider,
\begin{eqnarray}
   \rho_{xx}(t=0,z)&=& \frac{\rho_{\mu\mu}(t=0,z)+\rho_{\tau\tau}(t=0,z)}{2}\\
   \bar{\rho}_{xx}(t=0,z)&=& \frac{\bar{\rho}_{\mu\mu}(t=0,z)+\bar{\rho}_{\tau\tau}(t=0,z)}{2}
   \label{eq:prescription}
\end{eqnarray}
Since in a SN-like environment the initial flavor content for the nonelectron flavor neutrinos are approximately equal, the above prescription allows one to compare flavor evolution in the two cases with \emph{almost} similar initial conditions.

A remark is in order. The choice of initial conditions reported above is different from both what we assumed in our previous work \cite{Capozzi:2020kge} and from what is considered in Ref. \cite{Shalgar:2021wlj}. Let us first consider the former case. Here, the two approaches had the same initial flavor content for $\nu_e$ and $\bar{\nu}_e$. However, the three species approach was used assuming no $\nu_x$ at $t=0$, and the six species one had the same initial conditions presented in Figure \ref{fig:angdist}. Such assumptions introduced a difference in the total number of neutrinos between the two approaches.
In Ref. \cite{Shalgar:2021wlj}, the $e$ and $\mu$ flavor content was taken to be the same among the two approaches, but it was imposed $\rho_{\tau\tau}(t=0,z)=\bar{\rho}_{\tau\tau}(t=0,z)=0$. Here, despite having the same number of particles, setting one diagonal entry of the density matrix completely empty can intrinsically enhance the amount of flavor conversions.

 Figure \ref{fig:AllCase_Pee} shows the survival probability $P_{ee}$ as a function of time and space for the three species case (left panel) and the six species one (right panel). The qualitative natures of the solutions are similar, but we find some differences in the flavor outcome in the two scenarios. Considering $t>25$ ns, in the former case, $P_{ee}\sim 0.5$, while in the latter case, $P_{ee}\sim 0.4$. We emphasize again that a proper quantitative comparison between two- and three-flavor evolution should be performed in a way such that the total numbers of neutrinos are similar in the two cases. Otherwise, if one starts with a scenario where one of the nonelectron flavors, say $\nu_\tau$ and $\bar\nu_{\tau}$ for example, has a negligible population to start with, a large flavor conversion can be obtained simply because of the lack of entries in the density matrix. This might lead to an enhancement of the differences.  

\section{Conclusions}
\label{sec:Conclusions}

The evolution of neutrino fast flavor conversions depends on the occurrence of crossings in the angular distributions of flavor lepton number. In the context of supernova neutrinos, it is usually assumed that the flavor content of $\nu_\mu$, $\nu_\tau$, $\bar{\nu}_\mu$ and $\bar{\nu}_\tau$ is equal. Consequently, only three neutrino species are used ($\nu_e$, $\bar{\nu}_e$ and $\nu_x$) in numerical simulations, as well as linear stability analyses. Recently, the first hydrodynamical simulations with six neutrino species \cite{Bollig:2017lki} were performed. Driven by these, it has been pointed out that the differences expected between $\nu_\mu$ and $\bar{\nu}_\mu$, which are induced by a non-negligible population of negatively charged muons in the core, can introduce some observable modifications of the angular distributions of lepton number \cite{Capozzi:2020kge}. As a result, those angular crossings that are expected to occur with three neutrino species can be either erased or actually created. Furthermore, even assuming the same flavor content in both the three and six neutrino species approaches, a significant difference in the survival probabilities is observed \cite{Shalgar:2021wlj}. However, these conclusions have been obtained considering  a spatially homogeneous neutrino system. In this work we have relaxed this assumption and we have performed a numerical calculation in both space and time with six neutrino species.

We have considered the same four cases that we adopted in our previous work \cite{Capozzi:2020kge} and have solved the complete nonlinear equations of motion in one space and time dimension for each of them. It is evident from Fig.\,\ref{fig:AllCase} that out of the four cases considered, case 1 does not show any instability as there is no crossing in any of the sectors ($e-\mu$, $e-\tau$, and $\mu-\tau$), while the other three show absolute instabilities. Cases 2 and 4 have instabilities in all the three sectors whereas it is present only in the $\mu-\tau$ sector of case 3. The presence of absolute instability is evident from the fact that in all the cases the instability spreads around the point of origin without drifting. Moreover, the instability propagates in both directions around the origin point (upward and downward). This is because of the backward velocity modes ($v < 0$; see Figure \ref{fig:angdist}), which are present in all our numerical examples. The results obtained from the nonlinear analysis are confirmed by the linear one. However, we point out that the triggering of instabilities in the otherwise stable sector $\rho_{\alpha\beta}$, because of the absence of a crossing at $t=0$, can only be observed at the nonlinear level. Overall, we find that the qualitative nature of the results obtained in Ref.~\cite{Capozzi:2020kge} remains robust even when considering flavor evolution in both space and time. 

Overall, we find that considering all the six neutrino species can significantly affect the results obtained with only three of them. First, systems that are stable in the standard three species approach can be unstable in the more realistic six species one. Moreover, to make a fair comparison between the two approaches, we have done a numerical analysis starting with same flavor content for both cases, finding a relatively large difference in the flavor outcomes. Thus, this analysis emphasizes the need to include muons in the study of fast flavor conversions and in turn may reveal their influence on the supernova dynamics. 

\section{Acknowledgements}

The  work of F.C. is supported by GVA Grant No.CDEIGENT/2020/003. S.C. acknowledges the support of the Max Planck India Mobility Grant from the Max Planck Society, supporting the visit and stay at MPP during the project. S.C has also received funding from DST/SERB projects CRG/2021/002961 and MTR/2021/000540.

\bibliographystyle{JHEP}
\bibliography{biblio}

\end{document}